\documentclass{article}
\usepackage[T2A]{fontenc}
\usepackage[utf8]{inputenc}
\usepackage[english]{babel}
\usepackage{amsmath}
\usepackage{amsfonts}
\usepackage{amssymb}
\usepackage{wasysym}
\usepackage{amsthm}
\usepackage{tikz}
\usepackage{tikz-cd}
\usetikzlibrary{decorations.markings}
\usetikzlibrary{positioning}
\usepackage{array}
\usepackage{ytableau}
\usepackage{longtable}
\usepackage{empheq}
\usepackage{multicol}
\usepackage{mathrsfs}
\usepackage{amssymb}
\usepackage{diagbox}
\usepackage{physics}
\usepackage{pb-diagram}
\usepackage{setspace}
\usepackage{faktor}
\usepackage{indentfirst}
\usepackage{csquotes}
\usepackage[hyphens]{url}
\usepackage{breakurl}
\usepackage[hidelinks,colorlinks=true,unicode,breaklinks]{hyperref}
\hypersetup{linkcolor=blue, citecolor=blue, filecolor=blue, urlcolor=blue}
\usepackage[left=2.3cm,right=2.3cm,top=2.3cm,bottom=2.3cm,bindingoffset=0cm]{geometry}

\usepackage[natbib=true, backend = bibtex, style=numeric-comp, sorting=none]{biblatex}
\begin{document}
\title{{\Large {\bf Irreducible representations of simple Lie algebras\\ by differential operators}
\vspace{.2cm}}
\author{
{\bf A.~Morozov $^{c,a,b}$}\thanks{morozov.itep@mail.ru},
{\bf M.~Reva $^{c,a}$}\thanks{reva.ma@phystech.edu},
{\bf N.~Tselousov $^{c,a}$}\thanks{tselousov.ns@phystech.edu},
{\bf Y.~Zenkevich $^{e,f,g,a,d}$}\thanks{yegor.zenkevich@gmail.com} \date{ }}
}
\maketitle

\vspace{-6cm}

\begin{center}
	\hfill ITEP-TH-15/21\\
	\hfill IITP-TH-12/21\\
	\hfill MIPT-TH-11/21
\end{center}

\vspace{3.5cm}

\begin{center}

\begin{small}
$^a$ {\it ITEP, Moscow 117218, Russia}\\
$^b$ {\it IITP, Moscow 127994, Russia}\\
$^c$ {\it MIPT, Dolgoprudny, 141701, Russia}\\
$^d$ {\it ITMP, Moscow 119991, Russia}\\
$^e$ {\it SISSA, Triest, Italy}\\
$^f$ {\it INFN, Sezione di Trieste}\\
$^g$ {\it IGAP, Trieste, Italy}\\
\end{small}

\end{center}

\vspace{1cm}
\begin{abstract}
We describe a systematic method to construct arbitrary 
highest-weight modules, including arbitrary finite-dimensional representations, 
for any finite dimensional simple Lie algebra $\mathfrak{g}$.
The Lie algebra generators are represented as first order differential operators in
$\frac{1}{2} \left(\dim \mathfrak{g} - \text{rank} \, \mathfrak{g}\right) $ variables.
All rising generators ${\bf e}$ are universal in the sense that they do not depend on representation,
the weights enter (in a very simple way) only in the expressions for the lowering operators ${\bf f}$.
We present explicit formulas of this kind for the simple root generators of all classical Lie algebras.
\end{abstract}

\section{Introduction}
Groups and algebras play a distinguished role in modern physics, since they describe symmetries. 
Lie groups and algebras \cite{Goto, Bourbaki} are of special importance, since they underlie the structure of
fundamental physics in the framework of the Standard model \cite{Okun2}. 
However, the progress at the end of the previous century reveals that the true symmetries 
of the stringy models involve also affine \cite{Kac} and double affine algebras \cite{DingIohara, Miki, Awata2018svb, Awata2017lqa,Awata2017cnz,Awata2016riz,Awata2016mxc,Awata2016bdm}, 
which are clever generalization of the simple Lie algebras. 
The most straightforward approach to them is in terms of vertex operators \cite{BPZ,moonshine},
which realize particular representations through the relevantly defined/deformed
creation and annihilation operators --- also known as ``free fields''.
The main goal of this paper is to give an exhaustive description of this construction
in the case of ordinary Lie algebras --- what will then allow to reformulate
various questions (like the algebra of representations, Weyl reflections, 
duality/triality transfromations etc.)
in these terms, and discuss more interesting generalizations to other algebras
on this solid basis.
The main difference from the earlier considerations like in \cite{GKMMMO}
will be generic formulas for all simple root generators in all finite-dimensional
representation for the four classical series $A$, $B$, $C$ and $D$.
They represent the generators as first-order differential operators, depending  
on $\frac{1}{2} \left(\dim \mathfrak{g} - \text{rank} \, \mathfrak{g}\right) $ variables ---
it is the case when {\it all} the Casimir operators reduce to numbers
(less variables are not sufficient to describe {\it all} finite-dimensional
representations, while for more variables  the Casimirs become fully fledged 
differential operators).
In affine case the variables and differential operators are promoted to chiral free fields,
and are used in the free-field realization of Kac-Moody algebras in \cite{GMMOS,FeFr}.
In the optimal case the $c$-number Casimirs turn into screening operators \cite{DF,MMShakirov},
which makes the free-field description of conformal theories \cite{BPZ} more involved and interesting.
We will provide a more detailed description of affine and other generalizations elsewhere.

All the main features of these monomial representations can be seen from the simplest example of the following well-known highest-weight representation of $\mathfrak{sl}_2$:
\begin{align}
\label{sl2 monom}
    \begin{aligned}
    \textbf{e} &= \frac{\partial}{\partial X} \\
    \textbf{h} &= \lambda  - 2 X \frac{\partial}{\partial X} \\
    \textbf{f} &= \lambda X - X^2 \frac{\partial}{\partial X}
    \end{aligned}
\end{align}

Note, that $\textbf{h}\, \text{and}\, \textbf{f}$ generators have a polynomial part and a part that contains a differential operator.
The irreducible monomial representations have the following nice properties:
\begin{itemize}
     \item The monomial representation describes {\it all} finite dimensional irreducible highest weight representations in a simple and unified way. In the particular example \eqref{sl2 monom}, for $\lambda \in \mathbb{Z}_{\geqslant 0}$ one gets irreducible $\left(\lambda + 1\right)$-dimensional representation $V_{\lambda}$ with basis $\left(1, X,\ldots, X^{\lambda} \right)$ and $\lambda$ is actually the Dynkin label. However, in general case, $\lambda \in \mathbb{R}$ corresponds to infinite dimensional Verma module with basis $\left(1, X, X^2, X^3, \ldots \right)$;
    \item Lie algebra generators in the monomial representation have simple form: the first order differential operators with polynomial coefficients in $\frac{1}{2} \left(\dim \mathfrak{g} - \text{rank} \, \mathfrak{g} \right)$ variable. In the particular example of $\mathfrak{sl}_2$ one variable is used;
    \item The Casimir operators are {\it numbers} in these monomial representations, as it should be in the irreducible representations. In the case of $\mathfrak{sl}_2$ there is only one Casimir operator:
    \begin{equation}
        \textbf{C}_2 = \textbf{e} \, \textbf{f} + \textbf{f} \, \textbf{e} + \frac{1}{2} \textbf{h}^2 = \lambda + \frac{\lambda^2}{2}
    \end{equation}
    \item The raising operators ($\textbf{e}$ in our notation) do not depend on the highest weight $\lambda$ of the representation. In other words, these operators have {\it the same form} in all irreducible representations of the algebra $\mathfrak{g}$ and the corresponding formulas do not contain the Dynkin labels of the highest weight. The lowering operators ($\textbf{f}$ respectively) depend rather simply on the highest weight: only the coefficient in front of the  polynomial part depends on it. As a simple consequence, the highest vector of any representation can be chosen in the following form:
    \begin{equation}
        \ket{0}_{\lambda} = 1
    \end{equation}
    \item One can compute characters using the formulas for the monomial representations. The answers are the {\it quantum} dimensions. For example we provide the character of the fundamental representation $\lambda = 1$ using \eqref{sl2 monom}:
    \begin{equation}
        \text{ch}_{V_{\lambda}}(\epsilon) =\Tr_{V_{\lambda}} \exp{ \epsilon \textbf{h} } = \left. \Tr_{V_{\lambda}} \exp{ \epsilon \left(\lambda - 2 X \frac{\partial}{\partial X} \right)} \right|_{\lambda = 1} = e^{\epsilon} + e^{-\epsilon}
    \end{equation}
    It is well known that the quantum dimensions for the algebra $\mathfrak{sl}_2$ are the symmetric Schur polynomials $s_{\lambda}(x_1,x_2)$ at the special point $x_1 = e^{\epsilon}, x_2 = e^{-\epsilon}$, where $\epsilon$ is proportional to the sum of positive roots.
    \item One can consider tensor product of two irreducible representations. We provide example of highest weights $\lambda_X$ and $\lambda_Y$:
    \begin{equation}
    \begin{aligned}
        \Delta(\textbf{e})&=\partial_X + \partial_Y\\
        \Delta(\textbf{h})&=\lambda_X  +\lambda_Y -2X\partial_X-2Y\partial_Y\\
        \Delta(\textbf{f})&=\lambda_X X +\lambda_Y Y-X^2\partial_X-Y^2\partial_Y
    \end{aligned}
    \end{equation}
    These operators acts on the space $V_{\lambda_X} \otimes V_{\lambda_Y}$. The highest vectors $\ket{0}$ are described by the following equation:
    \begin{equation}
        \Delta(\textbf{e}) \ket{0} =0
    \end{equation}
    The solutions corresponding to finite dimensional representations are $(X-Y)^k$, where  $ 0 \leqslant k \leqslant \text{min}(\lambda_X,\lambda_Y)$. For $k$ not obeying the restriction, one gets the Verma module. For example, in the case $\lambda_X=2,\, \lambda_Y=1$ one gets the representation $V_{\lambda_X} \otimes V_{\lambda_Y}=\langle 1, X, X^2, Y, X Y , X^2 Y \rangle$ the highest vectors are $1$ and $X-Y$.
\end{itemize}

\section{Simple Lie algebras} \label{SimpleLieAlg}
In this section we list basic facts from the theory of simple Lie algebras. Let $r$ be the rank of a simple Lie algebra $\mathfrak{g}$. We construct $\mathfrak{g}$ using the Chevalley basis consisting of the Cartan generators $\textbf{h}_i$, generators $\textbf{e}_i$ corresponding to positive simple roots, and those corresponding to negative simple roots, $\textbf{f}_i$, where $i=1,\ldots,r$. The commutation relations in the Chevalley basis have the following form:
\begin{align}
    \label{KM_rel}
    \begin{aligned}
         [\textbf{h}_i, \textbf{h}_j] &= 0 \\
         [\textbf{h}_i, \textbf{e}_j] &= \mathcal{A}_{ji} \textbf{e}_j \\
         [\textbf{h}_i, \textbf{f}_j] &= -\mathcal{A}_{ji} \textbf{f}_j \\
         [\textbf{e}_i, \textbf{f}_j] &= \delta_{ij} \textbf{h}_j \\
    \end{aligned}
\end{align}
where $\mathcal{A}_{ij}$ is the Cartan matrix, i.e.\ an integral $n\times n$ matrix such that $\mathcal{A}_{ii}=2$, $\mathcal{A}_{ij}<0$, $\mathcal{A}_{ij}=0 \Leftrightarrow \mathcal{A}_{ji}=0$ and $\mathcal{A}$ is positive definite. $\mathcal{A}_{ij}$ can be defined using the Killing form $\langle \cdot, \cdot \rangle$, which is ad-invariant symmetric bilinear form:
\begin{equation}
    \mathcal{A}_{ij}=2\frac{\langle \alpha_i, \alpha_j \rangle}{\langle \alpha_j, \alpha_j \rangle}
\end{equation}
where $\alpha_i$ are positive simple roots. \\
The remaining positive and negative roots  $\Delta_{\pm}$ are obtained by the adjoint action of the simple roots on themselves. To correctly reproduce the non-simple roots one should take into account the Serre relations:
\begin{align}
    \label{Serre_rel}
    \begin{aligned}
         \left[ \text{ad}_{\textbf{e}_i} \right]^{1-\mathcal{A}_{ji}} \textbf{e}_j &= 0 \\
         \left[ \text{ad}_{\textbf{f}_i} \right]^{1-\mathcal{A}_{ji}} \textbf{f}_j &= 0 \\
    \end{aligned}
\end{align}
Note that in our notation the commutation relations and the Serre relations involve transposed Cartan matrix. An irreducilbe finite dimensional representation is completely characterized by its highest weight $\lambda$. It is convenient to expand $\lambda$ in the basis of fundamental weights $\omega_i$ defined by:
\begin{equation}
    2\frac{\langle \omega_i, \alpha_j\rangle}{\langle \alpha_j,\alpha_j\rangle}=\delta_{ij}
\end{equation}
where $i,j = 1, \ldots, r$. The expansion coefficients in this basis are called Dynkin labels. The Dynkin labels of an arbitrary highest weight have the form:
\begin{equation}
    \lambda = (\lambda_1, \ldots, \lambda_r), \hspace{10mm} \lambda_i \in \mathbb{Z}_{\geqslant 0}.
\end{equation}
The Dynkin labels of the fundamental weights are $\omega_i = (0,\ldots,1_i,\ldots, 0)$, where $i=1,\ldots,r$. To describe the highest weights of the irreduclble representations we use Dynkin labels, so the formulas explicitely depend on $\lambda_i, \ i=1,\ldots, r$.
\\

\subsection{Explicit formulas for the classical Lie algebras}
In the following sections we present explicit formulas for the each series of the classical Lie algebras. We provide basic facts about their dimensions, ranks, the numbers of positive roots and Cartan matrices. The monomial representations are constructed via the variables $X_{i,j}$ with two indices. The number of such variables and the range of the indices $i$, $j$ are specific to each case. We provide formulas only for the simple roots, while the other elements of the algebra can be constructed through the adjoint action of the simple roots on themselves, which is described by~\eqref{KM_rel} and~\eqref{Serre_rel}. Note, that the formulas we get for different series are essentially similar.

One important point should be emphasized: whenever the upper summation limit is less than the lower one, then we define the sum to be {\it vanishing.}

\subsection{$\text{A}_n$ series} \label{ASeries}
This series makes sense for $n \geqslant 1$ and corresponds to classical Lie algebras $\mathfrak{sl}_{n+1}$:
\begin{align}
    \begin{aligned}
         \dim \mathfrak{sl}_{n+1} &= n^2 + 2 n \\
         \text{rank} \,\mathfrak{sl}_{n+1} &= n \\
         |\Delta_{+}| &= \frac{n(n+1)}{2}
    \end{aligned}
\end{align}
\begin{equation}
\label{A_n explicit}
\mathcal{A}_{\textbf{A}} = \left(
\begin{array}{ccccccc}
 2 & -1 & 0 & . & 0 & 0 & 0 \\
 -1 & 2 & -1 & . & 0 & 0 & 0 \\
 0 & -1 & 2 & . & 0 & 0 & 0 \\
 . & . & . & . & . & . & . \\
 0 & 0 & 0 & . & 2 & -1 & 0 \\
 0 & 0 & 0 & . & -1 & 2 & -1 \\
 0 & 0 & 0 & . & 0 & -1 & 2 \\
\end{array}
\right)
\end{equation}
The polynomial representations are written in terms of the variables $X_{i,j}$, where $n \geqslant i \geqslant j > 0$. The total number of variables is $\sum_{i=1}^{n} i = \frac{n(n+1)}{2} = |\Delta_{+}|$ .
The simple roots have the following form for $k = n-1,\ldots, 0$:
\begin{align}
\boxed{
    \begin{aligned}
         \textbf{e}_{n-k} &= \frac{\partial}{\partial X_{k + 1,1}} + \sum_{i = k + 2}^{n} X_{i,i-k-1}\, \frac{\partial}{\partial X_{i,i-k}} \\
         \textbf{f}_{n-k} &= X_{k+1,1} \left(\lambda_{n-k} + \sum_{i = 1}^{k} X_{k,i}\, \frac{\partial}{\partial X_{k,i}} - \sum_{i = 1}^{k + 1} X_{k+1,i}\, \frac{\partial}{\partial X_{k + 1,i}} \right) + \\
         &+\sum_{i = k + 2}^{n} X_{i,i - k}\, \frac{\partial}{\partial X_{i,i-k-1}} - \sum_{i=1}^{k} X_{k+1,i+1}\, \frac{\partial}{\partial X_{k,i}}\\
         \textbf{h}_{n-k} &= \lambda_{n-k} + \sum_{i = 1}^{k} X_{k,i} \frac{\partial}{\partial X_{k,i}} - \sum_{i = 1}^{k + 1} \left( 1 + \delta_{i,1} \right) X_{k + 1,i} \frac{\partial}{\partial X_{k + 1,i}} +\\ &+\sum_{i = k + 2}^{n} \left( X_{i,i-k-1} \frac{\partial}{\partial X_{i,i-k-1}} - X_{i,i-k} \frac{\partial}{\partial X_{i,i-k}} \right)
    \end{aligned}
} \label{An}
\end{align}

\subsection{$\text{B}_{n}$ series} \label{BSeries}
This series makes sense for $n \geqslant 2$ and corresponds to the classical Lie algebras $\mathfrak{so}_{2n+1}$:
\begin{align}
    \begin{aligned}
         \dim \mathfrak{so}_{2n+1} &= 2 n^2 + n \\
         \text{rank} \,\mathfrak{so}_{2n+1} &= n \\
         |\Delta_{+}| &= n^2
    \end{aligned}
\end{align}
\begin{equation}
\mathcal{A}_{\textbf{B}} = \left(
\begin{array}{ccccccc}
 2 & -1 & 0 & . & 0 & 0 & 0 \\
 -1 & 2 & -1 & . & 0 & 0 & 0 \\
 0 & -1 & 2 & . & 0 & 0 & 0 \\
 . & . & . & . & . & . & . \\
 0 & 0 & 0 & . & 2 & -1 & 0 \\
 0 & 0 & 0 & . & -1 & 2 & -2 \\
 0 & 0 & 0 & . & 0 & -1 & 2 \\
\end{array}
\right)
\end{equation}
The polynomial representations are written in terms of the variables $X_{i,j}$, where $n \geqslant i >0 $ and $2i - 1 \geqslant j > 0$. The total number of variables is $\sum_{i=1}^{n} 2i - 1 = n^2 = |\Delta_{+}|$ .
The simple roots have the following form for $k = n-1,\ldots, 0$:
\begin{align}
\label{Bn}
\boxed{
    \begin{aligned}
         \textbf{e}_{n-k} &= \frac{\partial}{\partial X_{k + 1,1}} +
         \sum_{i = k + 2}^{n} \left( X_{i,i-k-1}\, \frac{\partial}{\partial X_{i,i-k}} +
         X_{i,i+k}\, \frac{\partial}{\partial X_{i,i+k+1}}  \right) \\
         \textbf{f}_{n-k} &= X_{k+1,1} \left(\lambda_{n-k} +
         \sum_{i = 1}^{2k-1} X_{k,i}\, \frac{\partial}{\partial X_{k,i}} -
         \sum_{i = 1}^{2k + 1} X_{k+1,i}\, \frac{\partial}{\partial X_{k + 1,i}} \right) + \\
         &+\left(1 + \delta_{k,0}\right)
         \sum_{i = k + 2}^{n}  \left( X_{i,i-k}\, \frac{\partial}{\partial X_{i,i-k-1}} +
         X_{i,i+k+1}\, \frac{\partial}{\partial X_{i,i+k}}\right) +\\
         & + \left(1 - \delta_{k,0} \right) \sum_{i=1}^{k+1} \frac{(-1)^{i+1}}{\left(1 + \delta_{k+1,i}\right)} X_{k+1,i}\, X_{k + 1, 2(k+1) - i}\, \frac{\partial}{\partial X_{k+1,2k+1}} - \sum_{i=1}^{2k-1} X_{k+1,i+1}\, \frac{\partial}{\partial X_{k,i}}\\
         \textbf{h}_{n-k} &= \lambda_{n-k} + \sum_{i=1}^{2k - 1} X_{k,i} \frac{\partial}{\partial X_{k,i}} - \sum_{i = 1}^{2k + \delta_{k,0}} \left( 1 + \delta_{i,1}  \right) X_{k + 1,i} \frac{\partial}{\partial X_{k + 1,i}} + \\ 
         &+\left( 1 + \delta_{k,0} \right) \sum_{i = k + 2}^{n} \left( X_{i,i-k-1} \frac{\partial}{\partial X_{i,i-k-1}} - X_{i,i+k+1} \frac{\partial}{\partial X_{i,i+k+1}}\right) -\\ 
         &- \left( 1 - \delta_{k,0} \right) \sum_{i = k + 2}^{n} \left( X_{i,i-k} \frac{\partial}{\partial X_{i,i-k}} - X_{i,i+k} \frac{\partial}{\partial X_{i,i+k}}\right)
    \end{aligned}
}
\end{align}

\subsection{$\text{C}_n$ series} \label{CSeries}
This series makes sense for $n \geqslant 2$ and corresponds to the classical Lie algebras $\mathfrak{sp}_{2n}$:
\begin{align}
    \begin{aligned}
         \dim \mathfrak{sp}_{2n} &= 2 n^2 + n \\
         \text{rank} \,\mathfrak{sp}_{2n} &= n \\
         |\Delta_{+}| &= n^2
    \end{aligned}
\end{align}
\begin{equation}
\mathcal{A}_{\textbf{C}} = \left(
\begin{array}{ccccccc}
 2 & -1 & 0 & . & 0 & 0 & 0 \\
 -1 & 2 & -1 & . & 0 & 0 & 0 \\
 0 & -1 & 2 & . & 0 & 0 & 0 \\
 . & . & . & . & . & . & . \\
 0 & 0 & 0 & . & 2 & -1 & 0 \\
 0 & 0 & 0 & . & -1 & 2 & -1 \\
 0 & 0 & 0 & . & 0 & -2 & 2 \\
\end{array}
\right)
\end{equation}
The polynomial representations are written in terms of the variables $X_{i,j}$, where $n \geqslant i >0 $ and $2i - 1 \geqslant j > 0$. The total number of variables is $\sum_{i=1}^{n} 2i - 1 = n^2 = |\Delta_{+}|$ .
The simple roots have the following form for $k = n-1,\ldots, 0$:
\begin{align}
\label{Cn}
\boxed{
    \begin{aligned}
         \textbf{e}_{n-k} &= \frac{\partial}{\partial X_{k + 1,1}} + \left(1 - \delta_{k,0}\right) X_{k+1,2k}\, \frac{\partial}{\partial X_{k + 1,2k+1}} +\\
         &+\sum_{i = k + 2}^{n} \left(  X_{i,i-k-1}\, \frac{\partial}{\partial X_{i,i-k}} +
         \left(1 - \delta_{k,0}\right) X_{i,i+k-1}\, \frac{\partial}{\partial X_{i,i+k}} \right)\\
         \textbf{f}_{n-k} &= X_{k+1,1} \left(\lambda_{n-k} +
         \sum_{i = 1}^{2k-1} X_{k,i}\, \frac{\partial}{\partial X_{k,i}} -
         \sum_{i = 1}^{2k + 1} X_{k+1,i}\, \frac{\partial}{\partial X_{k + 1,i}} \right)
         +\left(1 - \delta_{k,0}\right) X_{k+1,2k+1}\, \frac{\partial}{\partial X_{k + 1,2k}}\\
         &+\sum_{i = k + 2}^{n} \left(  X_{i,i-k}\, \frac{\partial}{\partial X_{i,i-k-1}} +
         \left(1 - \delta_{k,0}\right) X_{i,i+k}\, \frac{\partial}{\partial X_{i,i+k-1}} \right) +\\
         & + \left(1 - \delta_{k,0} \right) \sum_{i=1}^{2k-1} (-1)^{i+1} X_{k,i}\, X_{k + 1, 2k - i}\, \frac{\partial}{\partial X_{k,2k-1}} - \sum_{i=1}^{2k-1} \left(1 + \delta_{2k-1,i} \right) X_{k+1,i+1}\, \frac{\partial}{\partial X_{k,i}}\\
         \textbf{h}_{n-k} &= \lambda_{n-k} + \sum_{i = 1}^{2k-1} \left( 1 + \delta_{2k-1,i} \right) X_{k,i} \frac{\partial}{\partial X_{k,i}} - \sum_{i = 1}^{2k + 1} \left( 1 + \delta_{i,1} \right) X_{k + 1, i} \frac{\partial}{\partial X_{k + 1,i}} + \\ 
         &+ \left( 1 - \delta_{k,0} \right) \sum_{i = k + 1}^{n} \left( X_{i,i + k -1} \frac{\partial}{\partial X_{i,i+k-1}} - X_{i,i+k} \frac{\partial}{\partial X_{i,i+k}}\right) + \\ &+\sum_{i = k + 2}^{n} \left( X_{i,i - k -1} \frac{\partial}{\partial X_{i,i-k-1}} - X_{i,i-k} \frac{\partial}{\partial X_{i,i-k}}\right) 
    \end{aligned}
}
\end{align}

\subsection{$\text{D}_n$ series} \label{DSeries}
This series makes sense for $n \geqslant 3$ and corresponds to the classical Lie algebras $\mathfrak{so}_{2n}$:
\begin{align}
    \begin{aligned}
         \dim \mathfrak{so}_{2n} &= 2 n^2 - n \\
         \text{rank} \,\mathfrak{so}_{2n} &= n \\
         |\Delta_{+}| &= n^2 - n
    \end{aligned}
\end{align}
\begin{equation}
\mathcal{A}_{\textbf{D}} = \left(
\begin{array}{ccccccc}
 2 & -1 & 0 & . & 0 & 0 & 0 \\
 -1 & 2 & -1 & . & 0 & 0 & 0 \\
 0 & -1 & 2 & . & 0 & 0 & 0 \\
 . & . & . & . & . & . & . \\
 0 & 0 & 0 & . & 2 & -1 & -1 \\
 0 & 0 & 0 & . & -1 & 2 & 0 \\
 0 & 0 & 0 & . & -1 & 0 & 2 \\
\end{array}
\right)
\end{equation}
The polynomial representations are written in terms of the variables $X_{i,j}$, where $n \geqslant i > 1 $ and $2(i - 1) \geqslant j > 0$. The total number of variables is $\sum_{i=2}^{n} 2(i - 1) = n^2 - n = |\Delta_{+}|$ .
The simple roots have the following form for $k = n-1,\ldots, 0$:
\begin{align}
\label{Dn}
\boxed{
\begin{aligned}
         \textbf{e}_{n-k} &= \frac{\partial}{\partial X_{k + 1+\delta_{k,0},1+\delta_{k,0}}} +
         \sum_{i = k + 2 + \delta_{k,0}}^{n}  \left(X_{i,i-k-1-\delta_{k,0}}\, \frac{\partial}{\partial X_{i,i-k}} +
         X_{i,i+k-1}\, \frac{\partial}{\partial X_{i,i+k+\delta_{k,0}}} \right) \\
         \textbf{f}_{n-k} &= X_{k+1+ \delta_{k,0},1+\delta_{k,0}}\, \left(\lambda_{n-k} +
         \sum_{i = 1}^{2(k-1)} X_{k,i}\, \frac{\partial}{\partial X_{k,i}} -
         \sum_{i = 1}^{2k-\delta_{k,1}} X_{k+1,i}\, \frac{\partial}{\partial X_{k + 1,i}} - \delta_{k,0} X_{2,2}\, \frac{\partial}{\partial X_{2,2}}\right) + \\
         &+\sum_{i = k + 2 + \delta_{k,0}}^{n}  \left( X_{i,i-k}\, \frac{\partial}{\partial X_{i,i-k-1-\delta_{k,0}}} +
         X_{i,i+k+\delta_{k,0}}\, \frac{\partial}{\partial X_{i,i+k-1}} \right) + \\
         & + \sum_{i=1}^{k-\delta_{k,1}} (-1)^{i+1} X_{k+1,i}\, X_{k + 1, 2k+1-i}\, \frac{\partial}{\partial X_{k+1,2k}} -\sum_{i=1}^{2(k-1)} X_{k+1,i+1}\, \frac{\partial}{\partial X_{k,i}}\\
         \textbf{h}_{n-k} &= \lambda_{n-k} + \sum_{i = 1}^{2(k - 1)} X_{k,i} \frac{\partial}{\partial X_{k,i}} - \sum_{i = 1}^{2k-1} \left(1 + \delta_{i,1} \right) X_{k + 1, i} \frac{\partial}{\partial X_{k + 1, i}} - 2 \delta_{k,0} X_{2,2} \frac{\partial}{\partial X_{2,2}} +\\
         &+ \sum_{i = k + 2 + \delta_{k,0}}^{n} \left( X_{i,i-k-1-\delta_{k,0}} \frac{\partial}{\partial X_{i,i-k-1-\delta_{k,0}}} -X_{i,i-k+\delta_{k,0}} \frac{\partial}{\partial X_{i,i-k+\delta_{k,0}}}\right) +\\
         &+ \sum_{i = k + 2 + \delta_{k,0}}^{n} \left(X_{i,i+k-1} \frac{\partial}{\partial X_{i,i+k-1}} - X_{i,i+k} \frac{\partial}{\partial X_{i,i+k}}\right)
\end{aligned}
}
\end{align}

\section{Universal approach to simple Lie algebras}
Except for explicit formulas, we provide general algorithm for constructing the irreducible highest weight representations of finite dimensional Lie algebras. We discuss the number of variables, the dependence of the representations on the highest weight and possible changes of variables.

\subsection{The algorithm for constructing representations}
\label{rep_const_alg}
In this section we introduce the algorithm to construct the representations for arbitrary simple Lie algebra $\mathfrak{g}$. This algorithm was used to obtain formulas for classical Lie algebras \eqref{An}, \eqref{Bn}, \eqref{Cn}, \eqref{Dn}. Although the formulas for the classical Lie algebras are written in terms of $X_{i,j}$ variables, here we use different notation for simplicity: variables $X_i$, where $i = 1,\ldots, |\Delta_+|$.  \\
A monomial representation of highest weight $\lambda$ is determined by a map $\rho_{\lambda}$ that sends states $\ket{v}$ to polynomials. Lie algebra generators are sent to the first order differential operators. Schematically these rules read:
\begin{align}
    &\rho_{\lambda} \left( \ket{ v } \right) = \text{Pol} \! \left( X_1, \ldots, X_{|\Delta_{+}|} \right) \\
    &\rho_{\lambda} \left( \mathfrak{g} \right) = \text{Pol} \! \left( X_1, \ldots, X_{|\Delta_{+}|} \right) + \sum_{i = 1}^{|\Delta_{+}|} \text{Pol}_i \! \left( X_1, \ldots, X_{|\Delta_{+}|} \right) \frac{\partial}{\partial X_i}
\end{align}
where $\text{Pol} \! \left( X_1, \ldots, X_{|\Delta_{+}|} \right)$ are polynomials. The notation $\rho_{\lambda}$ is used in this section for demonstrative purposes, although in the other places we omit the mapping $\rho_{\lambda}$. The algorithm allow one to derive the simple root generators in arbitrary representation, while the other generators of the Lie algebra can be obtained from the simple roots. To construct the monomial representation one goes through the following stages:
\begin{enumerate}
    \item The form of the simple root generators $\textbf{e}_i, \textbf{f}_i, \ i = 1, \ldots, \text{rank}\, \mathfrak{g}$ in the representation of the highest weight $\lambda = \left( \lambda_1, \ldots, \lambda_{\text{rank}\, \mathfrak{g}} \right)$ is fixed by the following ansatz:
    \begin{equation}
    \label{EFansatz}
        \begin{aligned}
             \rho_{\lambda} \left( \textbf{e}_i \right) &=\frac{\partial}{\partial X_i} + \sum_{j=1,\, j\not=i }^{| \Delta_+|} A_{ij} \! \left( X_1, \ldots, X_{|\Delta_{+}|} \right)\frac{\partial}{\partial X_j} \\
             \rho_{\lambda} \left( \textbf{f}_i \right) &=\lambda_i X_i - X_i^2\frac{\partial}{\partial X_i} + \sum_{j=1,\, j\not=i }^{| \Delta_+ |} B_{ij}\! \left( X_1, \ldots, X_{|\Delta_{+}|} \right) \frac{\partial}{\partial X_j}
        \end{aligned}
    \end{equation}
    One can see that the first distinguished part of these operators form an $\mathfrak{sl}_2$ subalgebra that is identical to \eqref{sl2 monom}. The main advantage of this ansatz is that the polynomials $A_{ij}, B_{ij}$ have the {\it same} form in all representations $\lambda$. The rising operators do not depend on representation, while the lowering operators depend on representation only through the first term \eqref{EFansatz}.
    \item On the second stage we impose conditions to fix a remaining freedom of the ansatz. The highest vectors $\ket{0}$ of the fundamental representations $\omega_k, \ k = 1, \ldots, \text{rank} \, \mathfrak{g} $ are mapped to 1:
    \begin{equation}
        \rho_{\omega_k} \! \left( \ket{0} \right) = 1
    \end{equation}
    For each value of $i = \text{rank}\, \mathfrak{g} + 1, \ldots, |\Delta_{+}|$ we set:
    \begin{equation}
    \label{var_distrib}
        \rho_{\omega_{\gamma(i)}} \! \left( \textbf{f}_i \ket{0} \right) = X_i
    \end{equation}
    Here the value $\gamma(i)$ is picked up according to the following rule: for any positive root $\textbf{e}_j$ $j = 1, \ldots, |\Delta_{+}|$ the value $ \rho_{\omega_{\gamma(i)}} \! \left(\textbf{e}_j \textbf{f}_i \ket{0} \right)$ is defined on previous steps. This rule is not necessary in general, however it significantly simplifies the construction. After this stage the representation is obtained uniquely.

    \item On this stage one can find the form of polynomials $A_{ij}$, by demanding the appropriate action of $\textbf{e}_i$ in the fundamental representations $\omega_j$.
    \item After reconstructing all $\textbf{e}_i$ the other states of the representation $\omega_j$ can be obtained recursively by acting with $\textbf{e}_i$ on states $\ket{v}$ provided that $\rho_{\omega_j} \! \left( \textbf{e}_i \! \ket{v} \right)$ is already known for all $i = 1, \ldots, |\Delta_{+}|$. In other words, $\rho_{\omega_j} \! \left( \ket{v} \right)$ can be determined from the following system of equations:
    \begin{equation}
       \begin{cases}
              \rho_{\omega_j} \! \left( \textbf{e}_1 \right) \rho_{\omega_j} \! \left( \ket{v} \right) = \rho_{\omega_j} \! \left( \textbf{e}_1 \! \ket{v} \right)\\
             \cdots\cdots\cdots  \\
             \cdots\cdots\cdots  \\
             \rho_{\omega_j} \! \left( \textbf{e}_{|\Delta_+|} \right) \rho_{\omega_j} \! \left( \ket{v} \right) = \rho_{\omega_j} \! \left( \textbf{e}_{|\Delta_+|} \! \ket{v} \right)\\
        \end{cases}
    \end{equation}
    The right hand sides of the equations can not be zero for all equations simultaneously. The previous stages guarantee that the $\rho_{\omega_j} \! \left(\textbf{e}_i \right)$ are linearly independent, so the solution is unique.
    \item After determining enough states on the previous stage, we can reconstruct $\textbf{f}_i$ by action between two known states $\rho_{\omega_j} \! \left( \ket{v} \right)$ and $\rho_{\omega_j} \! \left( \textbf{f}_i \ket{v} \right)$.
\end{enumerate}

\subsection{Examples
}
We provide an example of the construction presented above in the case of $\mathfrak{sl}_3$. 
For $\mathfrak{sl}_3$ the root system is:
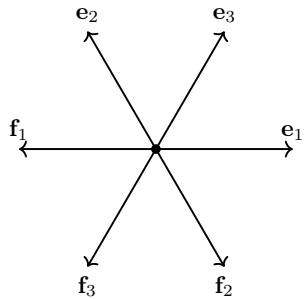
\begin{center}
\label{sl3 root system}
\resizebox{0.25\textwidth}{!}{%
    \begin{tikzpicture}
        \fill (0,0) circle (2pt);
        \draw[thick, ->] (0,0)  --   (2,0) node[above] {$\textbf{e}_1$};
        \draw[thick, ->] (0,0)  --   (1, 1.72) node[above] {$\textbf{e}_3$};
        \draw[thick, ->] (0,0)  --   (-1, 1.72) node[above] {$\textbf{e}_2$};
        \draw[thick, ->] (0,0)  --   (-2,0) node[above] {$\textbf{f}_1$};
        \draw[thick, ->] (0,0)  --   (-1, -1.72) node[below] {$\textbf{f}_3$};
        \draw[thick, ->] (0,0)  --   (1, -1.72) node[below] {$\textbf{f}_2$};
\end{tikzpicture}
}%
\end{center}
Lie algebra $\mathfrak{sl}_3$ has rank $r = 2$. Representations are described by two Dynkin labels $\left( \lambda_1, \lambda_2 \right)$. The algebra has three positive roots $\textbf{e}_1, \textbf{e}_2, \textbf{e}_3 = [\textbf{e}_1, \textbf{e}_2]$ and three negative roots $\textbf{f}_1, \textbf{f}_2, \textbf{f}_3 = [\textbf{f}_2, \textbf{f}_1]$ so we have to distribute $X_1, X_2, X_3$ among states of the fundamental representations $\omega_{1} = (1,0)$ and $\omega_2 = (0,1)$. To simplify formulas we omit the mapping $\rho$ and use the following notation for the highest vectors $\ket{0}_{i} := \rho_{\omega_i} (\ket{0})$.
The diagrams of the fundamental representations looks as follows:

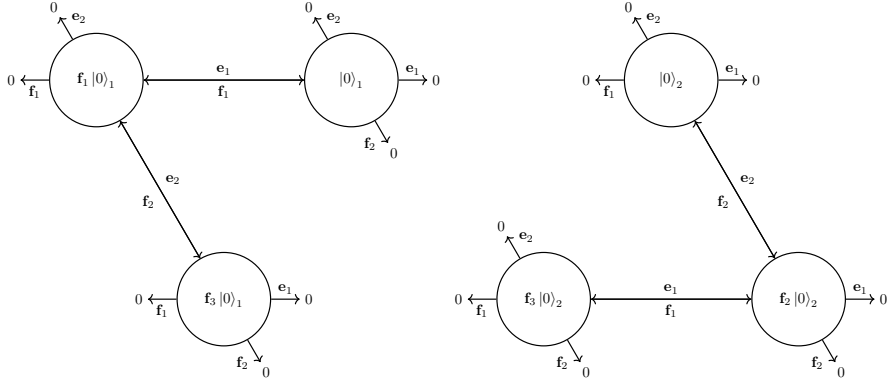
\begin{center}
\resizebox{0.35\textwidth}{!}{%
    \begin{tikzpicture}
        \node (a2) at (-5,0) {$0$};
        \node (a1) at (-3,0) [minimum size = 22mm, shape = circle, thick, draw] {$\textbf{f}_1 \ket{0}_1$};
        \node (a3) at (-4, 1.72) {$0$};

        \node (b1) at (3,0)  [minimum size = 22mm, shape = circle, thick, draw]  {$\ket{0}_1$};
        \node (b2) at (2,1.72) {$0$};
        \node (b3) at (4,-1.72) {$0$};
        \node (b4) at (5,0) {$0$};

        \node (c1) at (0, -5.16) [minimum size = 22mm, shape = circle, thick, draw] {$\textbf{f}_3 \ket{0}_1$};
        \node (c2) at (-2,-5.16) {$0$};
        \node (c3) at (2,-5.16) {$0$};
        \node (c4) at (1,-6.88) {$0$};

        \draw[thick, ->] (a1)  edge node[above right] {$\textbf{e}_2$} (a3);
        \draw[thick, ->] (c1)  edge node[above right] {$\textbf{e}_2$} (a1);
        \draw[thick, ->] (b1)  edge node[above right] {$\textbf{e}_2$} (b2);
        \draw[thick, ->] (a1)  edge node[above] {$\textbf{e}_1$} (b1);
        \draw[thick, ->] (b1)  edge node[above] {$\textbf{e}_1$} (b4);
        \draw[thick, ->] (c1)  edge node[above] {$\textbf{e}_1$} (c3);
        \draw[thick, ->] (a1)  edge node[below] {$\textbf{f}_1$} (a2);
        \draw[thick, ->] (b1)  edge node[below] {$\textbf{f}_1$} (a1);
        \draw[thick, ->] (c1)  edge node[below] {$\textbf{f}_1$} (c2);
        \draw[thick, ->] (c1)  edge node[below left] {$\textbf{f}_2$} (c4);
        \draw[thick, ->] (a1)  edge node[below left] {$\textbf{f}_2$} (c1);
        \draw[thick, ->] (b1)  edge node[below left] {$\textbf{f}_2$} (b3);
\end{tikzpicture}
}%
\resizebox{0.35\textwidth}{!}{%
    \begin{tikzpicture}
        \node (a2) at (-5,0) {$0$};
        \node (a1) at (-3,0) [minimum size = 22mm, shape = circle, thick, draw] {$\textbf{f}_3 \ket{0}_2$};
        \node (a3) at (-4, 1.72) {$0$};
        \node (a4) at (-2, -1.72) {$0$};

        \node (b1) at (3,0)  [minimum size = 22mm, shape = circle, thick, draw]  {$\textbf{f}_2 \ket{0}_2$};
        \node (b3) at (4,-1.72) {$0$};
        \node (b4) at (5,0) {$0$};

        \node (c1) at (0, 5.16) [minimum size = 22mm, shape = circle, thick, draw] {$\ket{0}_2$};
        \node (c2) at (-2,5.16) {$0$};
        \node (c3) at (2,5.16) {$0$};
        \node (c4) at (-1,6.88) {$0$};

        \draw[thick, ->] (a1)  edge node[above right] {$\textbf{e}_2$} (a3);
        \draw[thick, ->] (b1)  edge node[above right] {$\textbf{e}_2$} (c1);
        \draw[thick, ->] (a1)  edge node[above] {$\textbf{e}_1$} (b1);
        \draw[thick, ->] (b1)  edge node[above] {$\textbf{e}_1$} (b4);
        \draw[thick, ->] (c1)  edge node[above] {$\textbf{e}_1$} (c3);
        \draw[thick, ->] (a1)  edge node[below] {$\textbf{f}_1$} (a2);
        \draw[thick, ->] (b1)  edge node[below] {$\textbf{f}_1$} (a1);
        \draw[thick, ->] (c1)  edge node[below] {$\textbf{f}_1$} (c2);
        \draw[thick, ->] (c1)  edge node[above right] {$\textbf{e}_2$} (c4);
        \draw[thick, ->] (c1)  edge node[below left] {$\textbf{f}_2$} (b1);
        \draw[thick, ->] (b1)  edge node[below left] {$\textbf{f}_2$} (b3);
        \draw[thick, ->] (a1)  edge node[below left] {$\textbf{f}_2$} (a4);
\end{tikzpicture}
}%
\end{center}
\begin{enumerate}
    \item Two simple positive and negative roots have the form:
    \begin{align}
    \label{sl3 ansatz}
        \begin{aligned}
            \textbf{e}_1 &= \frac{\partial}{\partial X_1} + A_{12} \frac{\partial}{\partial X_2} + A_{13} \frac{\partial}{\partial X_3} \\
            \textbf{e}_2 &= A_{21} \frac{\partial}{\partial X_1} + \frac{\partial}{\partial X_2} + A_{23} \frac{\partial}{\partial X_3} \\
            \textbf{f}_1 &= \lambda_1 X_1 - X_1^2\frac{\partial}{\partial X_1} + B_{12} \frac{\partial}{\partial X_2} + B_{13} \frac{\partial}{\partial X_3} \\
            \textbf{f}_2 &= \lambda_2 X_2 + B_{21} \frac{\partial}{\partial X_1} - X_2^2 \frac{\partial}{\partial X_2}  + B_{23} \frac{\partial}{\partial X_3} \\
        \end{aligned}
    \end{align}
    To simplify notations we omit $X$ dependence of polynomials $A_{ij},B_{ij}$.
    \item The highest vectors are set to 1:
    \begin{align}
        \begin{aligned}
             \ket{0}_1 &= 1 \\
             \ket{0}_2 &= 1 \\
        \end{aligned}
    \end{align}
    There is one possible value $i = 3 = \text{rank} \, \mathfrak{sl}_3 + 1 = |\Delta_{+}|$. We have to set some state to be $X_3$. Possible yet undefined candidates are $\textbf{f}_3 \ket{0}_1$ or $\textbf{f}_3 \ket{0}_2$. They are both acceptable since all the states $\textbf{e}_i \textbf{f}_3 \ket{0}_j$ are already defined:
    \begin{align}
        \begin{aligned}
            \textbf{e}_1  \textbf{f}_3 \ket{0}_1  &= \textbf{f}_2 \ket{0}_1 = 0 &\hspace{20mm}  \textbf{e}_1  \textbf{f}_3 \ket{0}_2  &= \textbf{f}_2 \ket{0}_2 = X_2 \\
            \textbf{e}_2  \textbf{f}_3 \ket{0}_1  &= -\textbf{f}_1 \ket{0}_1 = -X_1 &\hspace{20mm}  \textbf{e}_2 \textbf{f}_3 \ket{0}_2 &= -\textbf{f}_1 \ket{0}_2 = 0 \\
            \textbf{e}_3  \textbf{f}_3 \ket{0}_1  &= \ket{0}_1 = 1 &\hspace{20mm}  \textbf{e}_3  \textbf{f}_3 \ket{0}_2  &= - \ket{0}_2 = -1 \\
        \end{aligned}
    \end{align}
    Note that one should be careful with the value of $\lambda$ in formulas \eqref{sl3 ansatz}. One can choose $\textbf{f}_3 \ket{0}_2 = X_3$, while $\textbf{f}_3 \ket{0}_1$ is determined uniquely at the following stages of the algorithm. On this stage the diagrams of the representations read:
    \begin{center}
\resizebox{0.35\textwidth}{!}
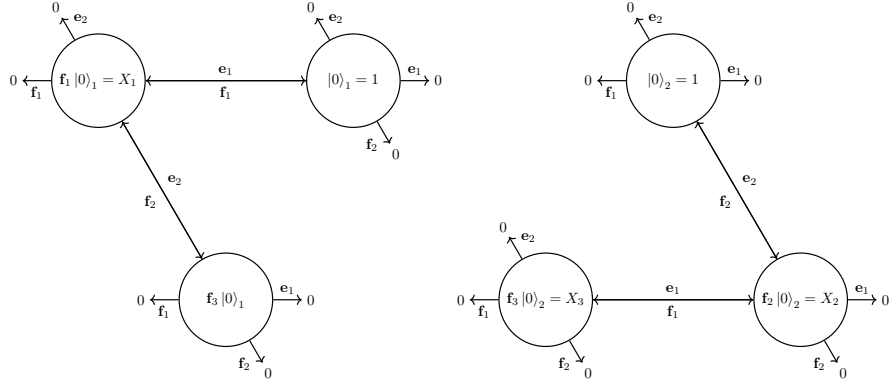
{%
    \begin{tikzpicture}
        \node (a2) at (-5,0) {$0$};
        \node (a1) at (-3,0) [minimum size = 22mm, shape = circle, thick, draw] {$\textbf{f}_1 \ket{0}_1 = X_1$};
        \node (a3) at (-4, 1.72) {$0$};

        \node (b1) at (3,0)  [minimum size = 22mm, shape = circle, thick, draw]  {$\ket{0}_1 = 1$};
        \node (b2) at (2,1.72) {$0$};
        \node (b3) at (4,-1.72) {$0$};
        \node (b4) at (5,0) {$0$};

        \node (c1) at (0, -5.16) [minimum size = 22mm, shape = circle, thick, draw] {$\textbf{f}_3 \ket{0}_1$};
        \node (c2) at (-2,-5.16) {$0$};
        \node (c3) at (2,-5.16) {$0$};
        \node (c4) at (1,-6.88) {$0$};

        \draw[thick, ->] (a1)  edge node[above right] {$\textbf{e}_2$} (a3);
        \draw[thick, ->] (c1)  edge node[above right] {$\textbf{e}_2$} (a1);
        \draw[thick, ->] (b1)  edge node[above right] {$\textbf{e}_2$} (b2);
        \draw[thick, ->] (a1)  edge node[above] {$\textbf{e}_1$} (b1);
        \draw[thick, ->] (b1)  edge node[above] {$\textbf{e}_1$} (b4);
        \draw[thick, ->] (c1)  edge node[above] {$\textbf{e}_1$} (c3);
        \draw[thick, ->] (a1)  edge node[below] {$\textbf{f}_1$} (a2);
        \draw[thick, ->] (b1)  edge node[below] {$\textbf{f}_1$} (a1);
        \draw[thick, ->] (c1)  edge node[below] {$\textbf{f}_1$} (c2);
        \draw[thick, ->] (c1)  edge node[below left] {$\textbf{f}_2$} (c4);
        \draw[thick, ->] (a1)  edge node[below left] {$\textbf{f}_2$} (c1);
        \draw[thick, ->] (b1)  edge node[below left] {$\textbf{f}_2$} (b3);
\end{tikzpicture}
}%
\resizebox{0.35\textwidth}{!}{%
    \begin{tikzpicture}
        \node (a2) at (-5,0) {$0$};
        \node (a1) at (-3,0) [minimum size = 22mm, shape = circle, thick, draw] {$\textbf{f}_3 \ket{0}_2 = X_3$};
        \node (a3) at (-4, 1.72) {$0$};
        \node (a4) at (-2, -1.72) {$0$};

        \node (b1) at (3,0)  [minimum size = 22mm, shape = circle, thick, draw]  {$\textbf{f}_2 \ket{0}_2 = X_2$};
        \node (b3) at (4,-1.72) {$0$};
        \node (b4) at (5,0) {$0$};

        \node (c1) at (0, 5.16) [minimum size = 22mm, shape = circle, thick, draw] {$\ket{0}_2 = 1$};
        \node (c2) at (-2,5.16) {$0$};
        \node (c3) at (2,5.16) {$0$};
        \node (c4) at (-1,6.88) {$0$};

        \draw[thick, ->] (a1)  edge node[above right] {$\textbf{e}_2$} (a3);
        \draw[thick, ->] (b1)  edge node[above right] {$\textbf{e}_2$} (c1);
        \draw[thick, ->] (a1)  edge node[above] {$\textbf{e}_1$} (b1);
        \draw[thick, ->] (b1)  edge node[above] {$\textbf{e}_1$} (b4);
        \draw[thick, ->] (c1)  edge node[above] {$\textbf{e}_1$} (c3);
        \draw[thick, ->] (a1)  edge node[below] {$\textbf{f}_1$} (a2);
        \draw[thick, ->] (b1)  edge node[below] {$\textbf{f}_1$} (a1);
        \draw[thick, ->] (c1)  edge node[below] {$\textbf{f}_1$} (c2);
        \draw[thick, ->] (c1)  edge node[above right] {$\textbf{e}_2$} (c4);
        \draw[thick, ->] (c1)  edge node[below left] {$\textbf{f}_2$} (b1);
        \draw[thick, ->] (b1)  edge node[below left] {$\textbf{f}_2$} (b3);
        \draw[thick, ->] (a1)  edge node[below left] {$\textbf{f}_2$} (a4);
\end{tikzpicture}
}%
\end{center}
    \item To find the form of $\textbf{e}_1, \textbf{e}_2$ we use the actions of the corresponding operators in the fundamental representations. These actions are clear from the diagrams and we use the following conditions:
    \begin{align}
        \begin{aligned}
            \textbf{e}_1  \textbf{f}_2 \ket{0}_2  &= 0 \hspace{10mm} &\Rightarrow \hspace{10mm} A_{12}(X) &= 0\\
            \textbf{e}_1  \textbf{f}_3 \ket{0}_2  &= -\textbf{f}_2 \ket{0}_2 = -X_2 \hspace{10mm} &\Rightarrow \hspace{10mm} A_{13}(X) &= -X_2\\
            \textbf{e}_2  \textbf{f}_1 \ket{0}_1  &= 0 \hspace{10mm} &\Rightarrow \hspace{10mm} A_{21}(X) &= 0\\
            \textbf{e}_2 \textbf{f}_3 \ket{0}_2  &= 0 \hspace{10mm} &\Rightarrow \hspace{10mm} A_{23}(X) &= 0\\
        \end{aligned}
    \end{align}
    We have completely determined the positive simple roots:
    \begin{align}
        \begin{aligned}
            \textbf{e}_1 &= \frac{\partial}{\partial X_1}  -X_2 \frac{\partial}{\partial X_3} \\
            \textbf{e}_2 &= \frac{\partial}{\partial X_2} \\
        \end{aligned}
    \end{align}
    \item There is only one undetermined state $ \textbf{f}_3 \ket{0}_1$. To determine it we need to solve the following system of equations:
    \begin{align}
        \begin{aligned}
            \textbf{e}_1  \textbf{f}_3 \ket{0}_1  &= 0 &\hspace{25mm} \left( \frac{\partial}{\partial X_1}  -X_2 \frac{\partial}{\partial X_3} \right) \left( \textbf{f}_3 \ket{0}_1 \right) &= 0 \\
            \textbf{e}_2  \textbf{f}_3 \ket{0}_1  &= \textbf{f}_1 \ket{0}_1 = X_1 &\hspace{25mm} \left( \frac{\partial}{\partial X_2} \right) \left( \textbf{f}_3 \ket{0}_1 \right) &= X_1 \\
            \textbf{e}_3  \textbf{f}_3 \ket{0}_1  &= \ket{0}_1 = 1 &\hspace{25mm} \left( \frac{\partial}{\partial X_3} \right) \left( \textbf{f}_3 \ket{0}_1 \right) &= 1\\
        \end{aligned}
    \end{align}
    The answer to this system reads:
    \begin{equation}
        \textbf{f}_3 \ket{0}_1 = X_3 + X_1 X_2
    \end{equation}
    \item On the last step we find $\textbf{f}_1, \textbf{f}_2$. To do this we again use actions of these operators in fundamental representations. Note that, $\textbf{f}_1, \textbf{f}_2$ operators depend on $\lambda$:
    \begin{align}
        \begin{aligned}
            \left. \textbf{f}_1 \left( \textbf{f}_2 \ket{0}_2 \right) \right|_{\lambda = (0,1)} &= -X_3 \hspace{10mm} &\Rightarrow \hspace{10mm} B_{12}(X) &= -X_3\\
            \left. \textbf{f}_1 \left( \textbf{f}_3 \ket{0}_2 \right)\right|_{\lambda = (0,1)} &= 0 \hspace{10mm} &\Rightarrow \hspace{10mm} B_{13}(X) &= 0\\
            \left. \textbf{f}_2 \left( \textbf{f}_1 \ket{0}_1 \right) \right|_{\lambda = (1,0)} &= \textbf{f}_3 \ket{0}_1 = X_3 + X_1 X_2 \hspace{10mm} &\Rightarrow \hspace{10mm} B_{21}(X) & = X_3 + X_1 X_2 \\
            \left. \textbf{f}_2 \left( \textbf{f}_3 \ket{0}_2 \right) \right|_{\lambda = (0,1)} &= 0 \hspace{10mm} &\Rightarrow \hspace{10mm} B_{23}(X) &=  - X_2 X_3\\
        \end{aligned}
    \end{align}
    The resulting $\mathfrak{sl}_3$ operators corresponding to simple roots have the following form:
    \begin{align}
    \boxed{
        \begin{aligned}
            \textbf{e}_1 &= \frac{\partial}{\partial X_1}  -X_2 \frac{\partial}{\partial X_3} \\
            \textbf{e}_2 &= \frac{\partial}{\partial X_2} \\
            \textbf{f}_1 &= \lambda_1 X_1 - X_1^2 \frac{\partial }{\partial X_1} - X_3 \frac{\partial }{\partial X_2} \\
            \textbf{f}_2 &= \lambda_2 X_2 + \left( X_3 + X_1 X_2 \right) \frac{\partial }{\partial X_1} - X_2^2 \frac{\partial }{\partial X_2} - X_2 X_3 \frac{\partial}{\partial X_3} \\
        \end{aligned}
        }
    \end{align}
\end{enumerate}
We provide also the non-simple root operators. They can be obtained from the simple roots operators: $\textbf{e}_3 = [\textbf{e}_1, \textbf{e}_2], \  \textbf{f}_3 = [\textbf{f}_2, \textbf{f}_1]$. The explicit form of these operators reads:
\begin{align}
    \begin{aligned}
         \textbf{e}_3 &= \frac{\partial}{\partial X_3} \\
         \textbf{f}_3 &= \lambda_1 \left( X_3 + X_1 X_2 \right) + \lambda_2 X_3 - X_1 \left( X_1 X_2 + X_3 \right) \frac{\partial}{\partial X_1} - X_2 X_3\frac{\partial}{\partial X_2} - X_3^2 
         \frac{\partial}{\partial X_3}
    \end{aligned}
\end{align}
Non-simple root operators have more involved formulas and can be reconstructed from the simple ones, hence we do not provide non-simple roots.

We emphasize that the algorithm is applicable to any simple Lie algebra. For example we provide the generators of $\text{G}_2$ algebra:
\begin{equation}
\boxed{
        \begin{aligned}
             \textbf{e}_1&=\frac{\partial}{\partial X_1} + X_3 \frac{\partial}{\partial X_4} + X_4 \frac{\partial}{\partial X_5}\\
             \textbf{e}_2&=\frac{\partial}{\partial X_2} + X_1 \frac{\partial}{\partial X_3} + X_5 \frac{\partial}{\partial X_6}\\
             \textbf{f}_1&=\lambda_1 X_1-X_1^2\frac{\partial}{\partial X_1} -3\left(X_3 - X_2 X_1 \right) \frac{\partial}{\partial X_2} +  \left( 2 X_4 - X_1 X_3 \right) \frac{\partial}{\partial X_3} + \\
             &+ \left( 2 X_5 - X_4 X_1 \right) \frac{\partial}{\partial X_4} - X_1 X_5 \frac{\partial}{\partial X_5} + \left( \frac{1}{2} X_4^2 - X_5 X_3 \right) \frac{\partial}{\partial X_6}\\
             \textbf{f}_2&=\lambda_2 X_2-X_2^2\frac{\partial}{\partial X_2}+X_3\frac{\partial}{\partial X_1} + X_6 \frac{\partial}{\partial X_5}
        \end{aligned}
}
\end{equation}

\subsection{Comment on the number of variables}
Every simple Lie algebra $\mathfrak{g}$ has $\lvert \Delta_{+} \rvert$ $\mathfrak{sl}_2$ subalgebras. The highest weight representation can be considered as a representation of $\lvert \Delta_{+} \rvert$ $\mathfrak{sl}_2$ subalgebras, each of which has its own highest weight. From the general theory we know, that we only need to consider the highest weights of the $\mathfrak{sl}_2$ subalgebras corresponding to the simple roots, as the other highest weights can be reconstructed uniquely. Each representation of $\mathfrak{sl}_2$ has at least one variable, so the minimal number of variables needed for the irreducible highest weight representation of $\mathfrak{g}$ is $\lvert \Delta_{+} \rvert$: if we take less variables, the elements of the algebra will become linearly dependent. Moreover, we can't take more variables, as it will lead to the representation becoming reducible and the reduction to the irreducible one can be done by applying some subspace constraint. For example, in $\mathfrak{sl}_2$ case one can construct the following representation:
\begin{equation}
    \begin{aligned}
         \textbf{e}&=Y\partial_X\\
         \textbf{h}&=Y\partial_Y-X\partial_X\\
         \textbf{f}&=X\partial_Y
    \end{aligned}
    \label{xy}
\end{equation}
This representations is not an irreducible one and has an infinite number of invariant subspaces, i.e.\ homogeneous polynomials. The constraint reducing to one of the subspaces can be chosen as
\begin{equation}
    \begin{aligned}
         X\partial_X+Y\partial_Y &= \lambda\\
         Y = \text{const} &= 1
    \end{aligned}
    \label{Y=const}
\end{equation}
In the second line of Eqs.~\eqref{Y=const} one can choose any constant, but for convenience we choose it to be 1. By solving Eqs.~\eqref{Y=const} one gets
\begin{equation}
    \begin{aligned}
    \partial_Y&=\lambda-X\partial_X\\
    \textbf{e}&=\partial_X\\
    \textbf{h}&=\lambda -2X\partial_X\\
    \textbf{f}&=\lambda X -X^2\partial_X
    \end{aligned}
\end{equation}
This representation is already irreducible, so further reduction is impossible.

\subsection{Changing the variables}
The representation we get is unique up to the diffeomorphic change of variables. This change can be performed as follows:
\begin{equation}
    \begin{aligned}
    X_i &= X_{i}(Y_1,\dots,Y_n)\\
    \frac{\partial}{\partial X_i}&=\sum_{j=1}^{n} \frac{\partial Y_{j}}{\partial X_i} \frac{\partial}{\partial Y_j}
    \end{aligned}
\end{equation}
For example, for $\mathfrak{sl}_3$ one of the possible representations is:
\begin{equation}
    \begin{aligned}
        \textbf{e}_1 &= \frac{\partial}{\partial X_1}  -X_2 \frac{\partial}{\partial X_3} \\
        \textbf{e}_2 &= \frac{\partial}{\partial X_2} \\
        \textbf{f}_1 &= \lambda_1 X_1 - X_1^2 \frac{\partial }{\partial X_1} - X_3 \frac{\partial}{\partial X_2} \\
        \textbf{f}_2 &= \lambda_2 X_2 + \left( X_3 + X_1 X_2 \right) \frac{\partial }{\partial X_1} - X_2^2 \frac{\partial }{\partial X_2} - X_2 X_3 \frac{\partial}{\partial X_3} \\
    \end{aligned}
\end{equation}
The corresponding fundamental representations are the following:
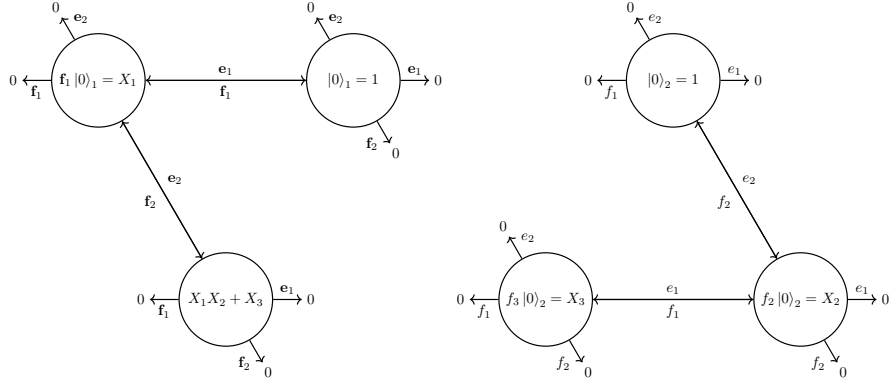
\begin{center}
\resizebox{0.35\textwidth}{!}{%
    \begin{tikzpicture}
        \node (a2) at (-5,0) {$0$};
        \node (a1) at (-3,0) [minimum size = 22mm, shape = circle, thick, draw] {$\textbf{f}_1 \ket{0}_1 = X_1$};
        \node (a3) at (-4, 1.72) {$0$};

        \node (b1) at (3,0)  [minimum size = 22mm, shape = circle, thick, draw]  {$\ket{0}_1 = 1$};
        \node (b2) at (2,1.72) {$0$};
        \node (b3) at (4,-1.72) {$0$};
        \node (b4) at (5,0) {$0$};

        \node (c1) at (0, -5.16) [minimum size = 22mm, shape = circle, thick, draw] {$X_1 X_2+X_3$};
        \node (c2) at (-2,-5.16) {$0$};
        \node (c3) at (2,-5.16) {$0$};
        \node (c4) at (1,-6.88) {$0$};

        \draw[thick, ->] (a1)  edge node[above right] {$\textbf{e}_2$} (a3);
        \draw[thick, ->] (c1)  edge node[above right] {$\textbf{e}_2$} (a1);
        \draw[thick, ->] (b1)  edge node[above right] {$\textbf{e}_2$} (b2);
        \draw[thick, ->] (a1)  edge node[above] {$\textbf{e}_1$} (b1);
        \draw[thick, ->] (b1)  edge node[above] {$\textbf{e}_1$} (b4);
        \draw[thick, ->] (c1)  edge node[above] {$\textbf{e}_1$} (c3);
        \draw[thick, ->] (a1)  edge node[below] {$\textbf{f}_1$} (a2);
        \draw[thick, ->] (b1)  edge node[below] {$\textbf{f}_1$} (a1);
        \draw[thick, ->] (c1)  edge node[below] {$\textbf{f}_1$} (c2);
        \draw[thick, ->] (c1)  edge node[below left] {$\textbf{f}_2$} (c4);
        \draw[thick, ->] (a1)  edge node[below left] {$\textbf{f}_2$} (c1);
        \draw[thick, ->] (b1)  edge node[below left] {$\textbf{f}_2$} (b3);
\end{tikzpicture}
}%
\resizebox{0.35\textwidth}{!}{%
    \begin{tikzpicture}
        \node (a2) at (-5,0) {$0$};
        \node (a1) at (-3,0) [minimum size = 22mm, shape = circle, thick, draw] {$f_3 \ket{0}_2 = X_3$};
        \node (a3) at (-4, 1.72) {$0$};
        \node (a4) at (-2, -1.72) {$0$};

        \node (b1) at (3,0)  [minimum size = 22mm, shape = circle, thick, draw]  {$f_2 \ket{0}_2 = X_2$};
        \node (b3) at (4,-1.72) {$0$};
        \node (b4) at (5,0) {$0$};

        \node (c1) at (0, 5.16) [minimum size = 22mm, shape = circle, thick, draw] {$\ket{0}_2 = 1$};
        \node (c2) at (-2,5.16) {$0$};
        \node (c3) at (2,5.16) {$0$};
        \node (c4) at (-1,6.88) {$0$};

        \draw[thick, ->] (a1)  edge node[above right] {$e_2$} (a3);
        \draw[thick, ->] (b1)  edge node[above right] {$e_2$} (c1);
        \draw[thick, ->] (a1)  edge node[above] {$e_1$} (b1);
        \draw[thick, ->] (b1)  edge node[above] {$e_1$} (b4);
        \draw[thick, ->] (c1)  edge node[above] {$e_1$} (c3);
        \draw[thick, ->] (a1)  edge node[below] {$f_1$} (a2);
        \draw[thick, ->] (b1)  edge node[below] {$f_1$} (a1);
        \draw[thick, ->] (c1)  edge node[below] {$f_1$} (c2);
        \draw[thick, ->] (c1)  edge node[above right] {$e_2$} (c4);
        \draw[thick, ->] (c1)  edge node[below left] {$f_2$} (b1);
        \draw[thick, ->] (b1)  edge node[below left] {$f_2$} (b3);
        \draw[thick, ->] (a1)  edge node[below left] {$f_2$} (a4);
\end{tikzpicture}
}%
\end{center}
One of the possible changes of variables is as follows:
\begin{equation}
    \begin{aligned}
         X_1&=Y_1 &\hspace{20mm} \frac{\partial}{\partial X_1}&=\frac{\partial}{\partial Y_1}+Y_2\frac{\partial}{\partial Y_3}\\
         X_2&= Y_2 &\hspace{20mm} \frac{\partial}{\partial X_2}&=\frac{\partial}{\partial Y_2}+Y_1\frac{\partial}{\partial Y_3}\\
         X_3&=Y_3-Y_1 Y_2 &\hspace{20mm} \frac{\partial}{\partial X_3}&=\frac{\partial}{\partial Y_3}
    \end{aligned}
\end{equation}
Applying this change of variables one gets the representation:
\begin{equation}
    \begin{aligned}
         \textbf{e}_1&=\frac{\partial}{\partial Y_1}\\
         \textbf{e}_2&=\frac{\partial}{\partial Y_2}+Y_1\frac{\partial}{\partial Y_3}\\
         \textbf{f}_1&=\lambda_1 Y_1-Y_1^2\frac{\partial}{\partial Y_1}+(Y_1 Y_2-Y_3)\frac{\partial}{\partial Y_2}-Y_1 Y_3\frac{\partial}{\partial Y_3}\\
         \textbf{f}_2&=\lambda_2 Y_2-Y_2^2\frac{\partial}{\partial Y_2}+Y_3\frac{\partial}{\partial Y_1}
    \end{aligned}
\end{equation}
After the change of variables the fundamental representations are:
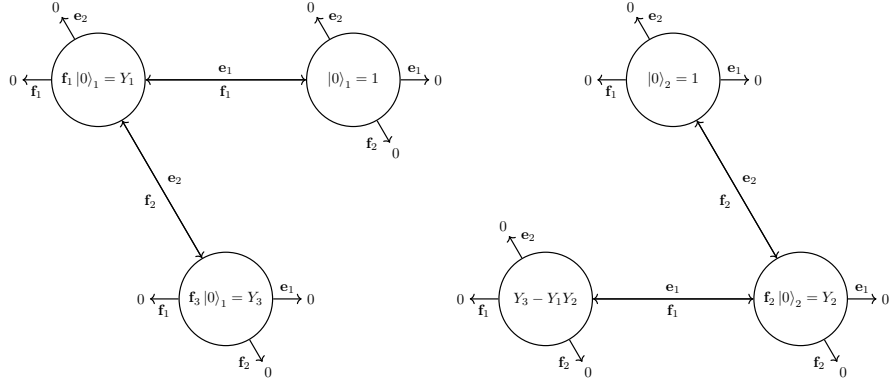
\begin{center}
\resizebox{0.35\textwidth}{!}{%
    \begin{tikzpicture}
        \node (a2) at (-5,0) {$0$};
        \node (a1) at (-3,0) [minimum size = 22mm, shape = circle, thick, draw] {$\textbf{f}_1 \ket{0}_1 = Y_1$};
        \node (a3) at (-4, 1.72) {$0$};

        \node (b1) at (3,0)  [minimum size = 22mm, shape = circle, thick, draw]  {$\ket{0}_1 = 1$};
        \node (b2) at (2,1.72) {$0$};
        \node (b3) at (4,-1.72) {$0$};
        \node (b4) at (5,0) {$0$};

        \node (c1) at (0, -5.16) [minimum size = 22mm, shape = circle, thick, draw] {$\textbf{f}_3\ket{0}_1=Y_3$};
        \node (c2) at (-2,-5.16) {$0$};
        \node (c3) at (2,-5.16) {$0$};
        \node (c4) at (1,-6.88) {$0$};

        \draw[thick, ->] (a1)  edge node[above right] {$\textbf{e}_2$} (a3);
        \draw[thick, ->] (c1)  edge node[above right] {$\textbf{e}_2$} (a1);
        \draw[thick, ->] (b1)  edge node[above right] {$\textbf{e}_2$} (b2);
        \draw[thick, ->] (a1)  edge node[above] {$\textbf{e}_1$} (b1);
        \draw[thick, ->] (b1)  edge node[above] {$\textbf{e}_1$} (b4);
        \draw[thick, ->] (c1)  edge node[above] {$\textbf{e}_1$} (c3);
        \draw[thick, ->] (a1)  edge node[below] {$\textbf{f}_1$} (a2);
        \draw[thick, ->] (b1)  edge node[below] {$\textbf{f}_1$} (a1);
        \draw[thick, ->] (c1)  edge node[below] {$\textbf{f}_1$} (c2);
        \draw[thick, ->] (c1)  edge node[below left] {$\textbf{f}_2$} (c4);
        \draw[thick, ->] (a1)  edge node[below left] {$\textbf{f}_2$} (c1);
        \draw[thick, ->] (b1)  edge node[below left] {$\textbf{f}_2$} (b3);
\end{tikzpicture}
}%
\resizebox{0.35\textwidth}{!}{%
    \begin{tikzpicture}
        \node (a2) at (-5,0) {$0$};
        \node (a1) at (-3,0) [minimum size = 22mm, shape = circle, thick, draw] {$Y_3-Y_1 Y_2$};
        \node (a3) at (-4, 1.72) {$0$};
        \node (a4) at (-2, -1.72) {$0$};

        \node (b1) at (3,0)  [minimum size = 22mm, shape = circle, thick, draw]  {$\textbf{f}_2 \ket{0}_2 = Y_2$};
        \node (b3) at (4,-1.72) {$0$};
        \node (b4) at (5,0) {$0$};

        \node (c1) at (0, 5.16) [minimum size = 22mm, shape = circle, thick, draw] {$\ket{0}_2 = 1$};
        \node (c2) at (-2,5.16) {$0$};
        \node (c3) at (2,5.16) {$0$};
        \node (c4) at (-1,6.88) {$0$};

        \draw[thick, ->] (a1)  edge node[above right] {$\textbf{e}_2$} (a3);
        \draw[thick, ->] (b1)  edge node[above right] {$\textbf{e}_2$} (c1);
        \draw[thick, ->] (a1)  edge node[above] {$\textbf{e}_1$} (b1);
        \draw[thick, ->] (b1)  edge node[above] {$\textbf{e}_1$} (b4);
        \draw[thick, ->] (c1)  edge node[above] {$\textbf{e}_1$} (c3);
        \draw[thick, ->] (a1)  edge node[below] {$\textbf{f}_1$} (a2);
        \draw[thick, ->] (b1)  edge node[below] {$\textbf{f}_1$} (a1);
        \draw[thick, ->] (c1)  edge node[below] {$\textbf{f}_1$} (c2);
        \draw[thick, ->] (c1)  edge node[above right] {$\textbf{e}_2$} (c4);
        \draw[thick, ->] (c1)  edge node[below left] {$\textbf{f}_2$} (b1);
        \draw[thick, ->] (b1)  edge node[below left] {$\textbf{f}_2$} (b3);
        \draw[thick, ->] (a1)  edge node[below left] {$\textbf{f}_2$} (a4);
\end{tikzpicture}
}%
\end{center}
Applying different changes of variables one can get from any distribution of variables on the diagrams of the fundamental representations to another one, but these distributions should give the consistent representations, i.e.\ the commutation relations in the algebra must be satisfied.

\subsection{Applying the universal approach to the classical Lie algebras}
In this section we discuss stable formulas \eqref{An}, \eqref{Bn}, \eqref{Cn}, \eqref{Dn} for the infinite series $A$, $B$, $C$ and $D$. By stability we understand the following property: the transition from $n$ to $n+1$ case involves only addition of new terms without a loss and changing of already existent ones.\\
To obtain stable formulas we choose specific form of function $\gamma(i)$ from the second stage of the algorithm \ref{rep_const_alg}. Let $\Delta_{+}^n$ be the set of its positive roots of a rank $n$ algebra of some type $A$, $B$, $C$, $D$. Then the analog of \eqref{var_distrib} reads:
    \begin{equation}
        \rho_{\omega_{k}} \left( \textbf{f}_i \ket{0} \right) = X_i, \hspace{10mm} i = |\Delta_{+}^{k-1}|, \ldots, |\Delta_{+}^{k}|
    \end{equation}
where $k = 1, \ldots, n$. Using this rule one obtains formulas \eqref{An}, \eqref{Bn}, \eqref{Cn}, \eqref{Dn}.
We argue that there are nice structures that arise after the renaming the variables $X_i \rightarrow X_{a,b}$. We provide explicit formulas for rising operators where sums along lines are implied.
\begin{itemize}
\item Rising operators/positive roots of $\text{A}_5$:
\begin{center}
    \begin{tikzpicture}
        \node (e1) at (-2,0) {$\textbf{e}_5=$};
        \node (e2) at (-2,1) {$\textbf{e}_4=$};
        \node (e3) at (-2,2) {$\textbf{e}_3=$};
        \node (e4) at (-2,3) {$\textbf{e}_2=$};
        \node (e5) at (-2,4) {$\textbf{e}_1=$};

        \node (a1) at (0,0) {$\partial_{1,1}$};

        \node (b1) at (2,1) {$\partial_{2,1}$};
        \node (b2) at (2,0) {$X_{2,1}\,\partial_{2,2}$};

        \node (c1) at (4,2) {$\partial_{3,1}$};
        \node (c2) at (4,1) {$X_{3,1}\,\partial_{3,2}$};
        \node (c3) at (4,0) {$X_{3,2}\,\partial_{3,3}$};

        \node (d1) at (6,3) {$\partial_{4,1}$};
        \node (d2) at (6,2) {$X_{4,1}\,\partial_{4,2}$};
        \node (d3) at (6,1) {$X_{4,2}\,\partial_{4,3}$};
        \node (d4) at (6,0) {$X_{4,3}\,\partial_{4,4}$};

        \node (g1) at (8,4) {$\partial_{5,1}$};
        \node (g2) at (8,3) {$X_{5,1}\,\partial_{5,2}$};
        \node (g3) at (8,2) {$X_{5,2}\,\partial_{5,3}$};
        \node (g4) at (8,1) {$X_{5,3}\,\,\partial_{5,4}$};
        \node (g5) at (8,0) {$X_{5,4}\,\partial_{5,5}$};

        \draw[->] (b1) edge (b2);
        \draw[->] (c1) edge (c2);
        \draw[->] (c2) edge (c3);
        \draw[->] (d1) edge (d2);
        \draw[->] (d2) edge (d3);
        \draw[->] (d3) edge (d4);
        \draw[->] (g1) edge (g2);
        \draw[->] (g2) edge (g3);
        \draw[->] (g3) edge (g4);
        \draw[->] (g4) edge (g5);
    \end{tikzpicture}
\end{center}
\item Rising operators/positive roots of $\text{B}_4$:
\begin{center}
    \begin{tikzpicture}
        \node (e1) at (-2,0) {$\textbf{e}_4=$};
        \node (e2) at (-2,1) {$\textbf{e}_3=$};
        \node (e3) at (-2,2) {$\textbf{e}_2=$};
        \node (e4) at (-2,3) {$\textbf{e}_1=$};

        \node (a1) at (0,0) {$\partial_{1,1}$};

        \node (b1) at (2,1) {$\partial_{2,1}$};
        \node (b2) at (2,0) {$X_{2,1}\,\partial_{2,2}$};
        \node (bl2) at (2,-0.5) {};
        \node (b3) at (4,0) {$X_{2,2}\,\partial_{2,3}$};
        \node (bl3) at (4,-0.5) {};

        \node (c1) at (6,2) {$\partial_{3,1}$};
        \node (c2) at (6,1) {$X_{3,1}\,\partial_{3,2}$};
        \node (c3) at (6,0) {$X_{3,2}\,\partial_{3,3}$};
        \node (cl3) at (6,-0.5) {};
        \node (c4) at (8,0) {$X_{3,3}\,\partial_{3,4}$};
        \node (cl4) at (8,-0.5) {};
        \node (c5) at (8,1) {$X_{3,4}\,\partial_{3,5}$};

        \node (d1) at (10,3) {$\partial_{4,1}$};
        \node (d2) at (10,2) {$X_{4,1}\,\partial_{4,2}$};
        \node (d3) at (10,1) {$X_{4,2}\,\partial_{4,3}$};
        \node (d4) at (10,0) {$X_{4,3}\,\partial_{4,4}$};
        \node (d5) at (12,0) {$X_{4,4}\,\partial_{4,5}$};
        \node (dl4) at (10,-0.5) {};
        \node (dl5) at (12,-0.5) {};
        \node (d6) at (12,1) {$X_{4,5}\,\partial_{4,6}$};
        \node (d7) at (12,2) {$X_{4,6}\,\partial_{4,7}$};

        \draw[->] (b1) edge (b2);
        \draw[->] (bl2) edge (bl3);

        \draw[->] (c1) edge (c2);
        \draw[->] (c2) edge (c3);
        \draw[->] (cl3) edge (cl4);
        \draw[->] (c4) edge (c5);

        \draw[->] (d1) edge (d2);
        \draw[->] (d2) edge (d3);
        \draw[->] (d3) edge (d4);
        \draw[->] (dl4) edge (dl5);
        \draw[->] (d5) edge (d6);
        \draw[->] (d6) edge (d7);

    \end{tikzpicture}
\end{center}
\item Rising operators/positive roots of $\text{C}_4$:
\begin{center}
    \begin{tikzpicture}
        \node (e1) at (-2,0) {$\textbf{e}_4=$};
        \node (e2) at (-2,1) {$\textbf{e}_3=$};
        \node (e3) at (-2,2) {$\textbf{e}_2=$};
        \node (e4) at (-2,3) {$\textbf{e}_1=$};

        \node (a1) at (0,0) {$\partial_{1,1}$};

        \node (b1) at (2,1) {$\partial_{2,1}$};
        \node (b2) at (2,0) {$X_{2,1}\,\partial_{2,2}$};
        \node (b3) at (4,1) {$X_{2,2}\,\partial_{2,3}$};

        \node (c1) at (6,2) {$\partial_{3,1}$};
        \node (c2) at (6,1) {$X_{3,1}\,\partial_{3,2}$};
        \node (c3) at (6,0) {$X_{3,2}\,\partial_{3,3}$};
        \node (c4) at (8,1) {$X_{3,3}\,\partial_{3,4}$};
        \node (c5) at (8,2) {$X_{3,4}\,\partial_{3,5}$};

        \node (d1) at (10,3) {$\partial_{4,1}$};
        \node (d2) at (10,2) {$X_{4,1}\,\partial_{4,2}$};
        \node (d3) at (10,1) {$X_{4,2}\,\partial_{4,3}$};
        \node (d4) at (10,0) {$X_{4,3}\,\partial_{4,4}$};
        \node (d5) at (12,1) {$X_{4,4}\,\partial_{4,5}$};
        \node (d6) at (12,2) {$X_{4,5}\,\partial_{4,6}$};
        \node (d7) at (12,3) {$X_{4,6}\,\partial_{4,7}$};

        \draw[->] (b1) edge (b2);
        \draw[->] (b2) edge (b3);

        \draw[->] (c1) edge (c2);
        \draw[->] (c2) edge (c3);
        \draw[->] (c3) edge (c4);
        \draw[->] (c4) edge (c5);

        \draw[->] (d1) edge (d2);
        \draw[->] (d2) edge (d3);
        \draw[->] (d3) edge (d4);
        \draw[->] (d4) edge (d5);
        \draw[->] (d5) edge (d6);
        \draw[->] (d6) edge (d7);

    \end{tikzpicture}
\end{center}
\item Rising operators/positive roots of $\text{D}_5$:
\begin{center}
    \begin{tikzpicture}
        \node (e1) at (-2,0) {$\textbf{e}_5=$};
        \node (e2) at (-2,1) {$\textbf{e}_4=$};
        \node (e3) at (-2,2) {$\textbf{e}_3=$};
        \node (e4) at (-2,3) {$\textbf{e}_2=$};
        \node (e5) at (-2,4) {$\textbf{e}_1=$};

        \node (a1) at (0,0) {$\partial_{2,2}$};
        \node (a2) at (0,1) {$\partial_{2,1}$};

        \node (b1) at (2,2) {$\partial_{3,1}$};
        \node (b2) at (2,1) {$X_{3,1}\,\partial_{3,2}$};
        \node (b3) at (2,0) {$X_{3,1}\,\partial_{3,3}$};
        \node (b4) at (4,1) {$X_{3,3}\, \partial_{3,4}$};
        \node (b5) at (4,0) {$X_{3,2}\, \partial_{3,4}$};

        \node (c1) at (6,3) {$\partial_{4,1}$};
        \node (c2) at (6,2) {$X_{4,1}\,\partial_{4,2}$};
        \node (c3) at (6,1) {$X_{4,2}\,\partial_{4,3}$};
        \node (c4) at (6,0) {$X_{4,2}\,\partial_{4,4}$};
        \node (c5) at (8,1) {$X_{4,4}\,\partial_{4,5}$};
        \node (c6) at (8,0) {$X_{4,3}\,\partial_{4,5}$};
        \node (c7) at (8,2) {$X_{4,5}\,\partial_{4,6}$};

        \node (d1) at (10,4) {$\partial_{5,1}$};
        \node (d2) at (10,3) {$X_{5,1}\,\partial_{5,2}$};
        \node (d3) at (10,2) {$X_{5,2}\,\partial_{5,3}$};
        \node (d4) at (10,1) {$X_{5,3}\,\partial_{5,4}$};
        \node (d5) at (10,0) {$X_{5,3}\,\partial_{5,5}$};
        \node (d6) at (12,1) {$X_{5,5}\,\partial_{5,6}$};
        \node (d7) at (12,0) {$X_{5,4}\,\partial_{5,6}$};
        \node (d8) at (12,2) {$X_{5,6}\,\partial_{5,7}$};
        \node (d9) at (12,3) {$X_{5,7}\,\partial_{5,8}$};

        \draw[->] (b1) edge (b2);
        \draw[-] (b2) edge (b3);
        \draw[->] (b3) edge (b4);
        \draw[->] (b2) edge (b5);
        \draw[-] (b5) edge (b4);

        \draw[->] (c1) edge (c2);
        \draw[->] (c2) edge (c3);
        \draw[-] (c3) edge (c4);
        \draw[-] (c5) edge (c6);
        \draw[->] (c4) edge (c5);
        \draw[->] (c3) edge (c6);
        \draw[->] (c5) edge (c7);

        \draw[->] (d1) edge (d2);
        \draw[->] (d2) edge (d3);
        \draw[->] (d3) edge (d4);
        \draw[-] (d4) edge (d5);
        \draw[-] (d6) edge (d7);
        \draw[->] (d4) edge (d7);
        \draw[->] (d5) edge (d6);
        \draw[->] (d6) edge (d8);
        \draw[->] (d8) edge (d9);
    \end{tikzpicture}
\end{center}
Note that there are strange-looking crossings in the last two operators $\textbf{e}_4, \textbf{e}_5$. It corresponds to the symmetry of the $\text{D}_n$ Dynkin diagrams --- the last two ``spinor'' roots can be swaped.
\end{itemize}

\section{Conclusion}
This paper provides an explicit realization of all finite-dimensional
representations of all classical Lie algebras $\mathfrak{g}$ in terms of
differential operators in 
$\frac{1}{2} \left(\dim \mathfrak{g} - \text{rank} \, \mathfrak{g}\right) $ variables.
Together with the weights these are the variables which get promoted to
$\dim \mathfrak{g}$ free fields in the affine  case,
and this work can serve as a basis for deriving the Kac-Moody bosonization formulas
in full generality (and not just in particular examples, as it was done in \cite{GMMOS}).
Other obvious directions include super, elliptic and DIM algebras.
Remarkable simplicity of our results in the classical case raises the hope
that these generalization can also be handled in a universal form for
large classes of representations.
This would open a way for systematic study of physical theories with DIM symmetry \cite{Awata2016bdm,Awata2016mxc,Awata2016riz,Awata2017cnz,Awata2017cnz,Awata2017lqa,Awata2018svb}
(like 5-brane networks \cite{Mironov:2016yue, Ghoneim2020sqi,Zenkevich2018fzl,Zenkevich2019ayk,Zenkevich2020ufs, Awata2016bdm} etc.).

Every raising generator ${\bf e}_{\vec\alpha}$, associated with the root $\vec\alpha$ has a leading term
$\partial/\partial X_{\vec\alpha}$ and a number of additional terms in which other derivatives
are multiplied by powers of $X$.
They are universal, i.e.\ do not depend on the weights of the representation.
The dependence on representation appears only in the lowering operators ${\bf f}_{\vec\alpha}$,
in the form of polynomials of $X$ multiplied by the Dynkin labels.
As a corollary, it shows up also in Cartan operators ${\bf h}$, just as $X$-independent terms.
Our formulas describe highest weight representations, which become also lowest weight
(i.e. finite-dimensional) only for non-negative integer-valued weights.
We found a general algorithm, which has a pronounced triangular structure but does not refer
explicitly to Gauss or Iwasava decompositions --- like it was done in \cite{GKMMMO}.
Instead we get a full and explicit description for all simple algebras,
which can be directly applied for practical purposes, not just to build an abstract theory. 
Also one can study what happens beyond the usual Dynkin diagrams, 
i.e.\ when the finite-growth restriction is raised, 
and analyze the problems like Vogel universality \cite{Vogel, Vogel2, Mironov:2015era, Mironov:2015ffv, Isaev:2021onl}, which arises in the ``adjoint sector''.
We do not go in these directions in the present paper in order to leave it clean and transparent:
classical series deserve a separate clear presentation. 

\section*{Acknowledgements}
Our work was partly supported by the grants of the Foundation for the Advancement of Theoretical Physics ``BASIS'' (A.M., N.T.), by RFBR grants 19-02-00815 (A.M., Y.Z.), 20-01-00644 (N.T.), by joint RFBR grants 19-51-18006-Bolg\_a (A.M., Y.Z.), 21-51-46010-CT\_a (A.M., N.T., Y.Z.), 21-52-52004-MOST (A.M., Y.Z.), .

\printbibliography

\end{document}